\documentclass[epj]{svjour}

\usepackage{graphicx}
\usepackage{fancyhdr}
\def\makeheadbox{{%  remove EPJC header
 }}
\def\mytitle{My title} 
\def\myauthors{My name}  
\def\mytype{My type of session}
\def\mysession{My session}
\def\mytitle{Radiative Yukawa couplings for SUSY Higgs
singlets at large tan(beta)} %Put your title here!
\def\myauthors{Robert N. Hodgkinson}    %Put your name here!
\def\mytype{Contributed Talk}    
\def\mysession{Colliders - Higgs Phenomenology}
\begin{document}
\title{Tan(beta) enhanced Yukawa couplings for supersymmetric Higgs singlets
at one loop}
\author{Robert N. Hodgkinson\inst{1}
%\thanks{\emph{Email:} Insert  Email  of corresponding author here}%
% \and
% Apostolos Pilaftsis\inst{1}% etc
}                     % Do not remove
\institute{School of Physics and Astronomy, University of Manchester,
Manchester M13 9PL, United Kingdom
%\and the second institute address here
}
\date{}
\abstract{
Extensions of the MSSM generically feature gauge singlet Higgs bosons.
These singlet Higgs bosons have $\tan\beta$-enhanced 
Yukawa couplings to down-type quarks and leptons at the one-loop level. 
We present an effective Lagrangian incorporating these Yukawa couplings
and use it to study their effect on singlet Higgs boson phenomenology 
within both the mnSSM and NMSSM. It is found that the loop-induced couplings
represent an appreciable effect for the singlet pseudoscalar in particular,
and may dominate its decay modes in some regions of parameter space.
 % end of PACS codes
} %end of abstract
\maketitle
%DO NOT REMOVE THIS LINE
%

\section{Introduction}
\label{intro}

The Minimal Supersymmetric Standard Model (MSSM) is a well-motivated
extension of the Standard Model of particle physics (SM), which provides
a technical solution to the gauge hierarchy problem. The model is minimal
in the sense that it includes only those terms in the superpotential
which are phenomenologically required, namely the Yukawa couplings
${\bf h}_u,{\bf h}_d,{\bf h}_e$ and a Higgs mass term $\mu$.

One theoretical weakness of the MSSM is the so-called $\mu$-problem
\cite{Kim:1983dt,CPNSH}. In order to achieve a successful electroweak
symmetry breaking scheme, the $\mu$-parameter describing the mixing
of the two Higgs superfields in the superpotential, i.e.
$\mu\hat H_u\hat H_d$, must be of the order of the soft SUSY-breaking
scale $M_{\rm SUSY}\sim 1$ TeV. Within the context of supergravity
(SUGRA), the $\mu$-parameter is not in general protected from gravity
effects, and is expected to be of the order the Planck scale
$M_{\rm Pl}$.

A natural solution to the $\mu$-problem may be obtained by extending
the MSSM to include a third Higgs superfield $\hat S$, which is a
singlet under the SM gauge group, and replacing the $\mu$-term in the
superpotential by $\lambda \hat S \hat H_u \hat H_d$. When supersymmetry
is softly broken, the scalar component $S$ of $\hat S$ generically
acquires a vacuum expectation value (VEV) of order $M_{\rm SUSY}$,
giving rise to an effective $\mu$-term of the required order.

\begin{figure}
\includegraphics[width=0.45\textwidth,height=0.22\textwidth,angle=0]
{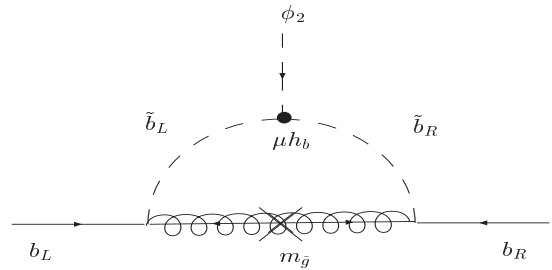}
\caption{\it The dominant contribution to the inhomogenious coupling
$\phi_2 b\bar b$ in the MSSM at large $\tan\beta$.}
\label{fig:1}       % Give a unique label
\end{figure}

The superpotential of such a singlet extension of the MSSM exhibits
an unwanted global Peccei-Quinn (PQ) symmetry $U(1)_{\rm PQ}$, unless
further additions or assumptions are made to the model. The PQ
symmetry must be explicitly broken above the electroweak scale to avoid
the appearance of visible axions after spontaneous symmetry breaking.
Several models have been proposed in the literature based on different
choices of discrete and gauged symmetries to break the PQ symmetry
\cite{CPNSH}, including the Next-to-Minimal Supersymmetric SM (NMSSM)
\cite{NMSSM}, the minimal nonminimal Supersymmetric SM (mnSSM)
\cite{Panagiotakopoulos:2000wp} and the $U(1)'$-extended
Supersymmetric SM (UMSSM)\cite{Cvetic:1997ky}.

A common feature of all these models is that the singlet Higgs boson
has no tree level couplings to SM fermions or gauge bosons. It has long
been known \cite{TBanks,Hempfling:1993kv} that within the MSSM, threshold
 corrections to the Yukawa couplings to $b$ quarks and $\tau$ leptons can
become significant in the limit of large $\tan\beta$, where $\tan\beta$
is the ratio of the two Higgs VEVs. This enhancement partially overcomes
the loop suppression factor, and in regions where mixing between the Higgs
particles is negligible, the one-loop correction can dominate the
$H_1\rightarrow b\bar b$ decay width \cite{Coarasa:1995yg}. The dominant
contribution to the inhomogenous coupling $\phi_2 b\bar b$ is shown in
Fig.~\ref{fig:1}.

\begin{figure}
\includegraphics[width=0.45\textwidth,height=0.22\textwidth,angle=0]
{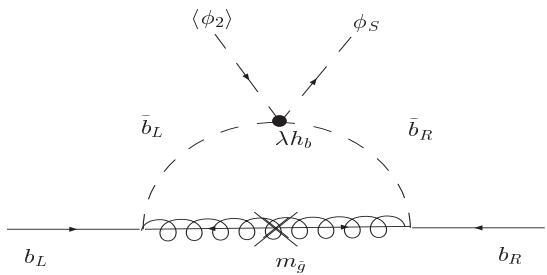}
\caption{\it The dominant contribution to the coupling $\phi_S b\bar b$ in
singlet extensions of the MSSM at large $\tan\beta$, analogous to the
MSSM graph of Fig~\ref{fig:1}.}
\label{fig:2}       % Give a unique label
\end{figure}

An analogous $\tan\beta$ enhanced Yukawa coupling for the singlet Higgs
boson is generated at one-loop through sfermion-gaugino loops in singlet
extensions of the MSSM \cite{Hodgkinson}.
The dominant contribution to the $\phi_S b\bar b$
coupling is shown in Fig.~\ref{fig:2}. These effective couplings can
be significant, e.g. of order the SM Yukawa couplings, and in the limit
where the $H_d$ doublet decouples from the low energy spectrum, they can
provide the dominant decay mechanism for light singlets.

\vspace{-0.25in}

\section{Effective Lagrangian framework}
\label{sec:1}

The general effective Lagrangian for the self-energy transition
$f_L\to f_R$ in the nonvanishing Higgs background may be written
as

\begin{equation}
-{\mathcal L}^f_{\rm self} = 
h_f \bar f_R\left(
\Phi_1^{0\ast}+\Delta_f\left[\Phi_1^0,\Phi_2^0,S\right]
\right)f_L + {\rm H.c.}
\end{equation}
where $\Phi_{1,2}^0={1\over \sqrt{2}}\left(v_{1,2}+\phi_{1,2}
+ia_{1,2}\right)$ are the electrically neutral components of the
two Higgs doublets $H_{d,u}$\footnote{Here we adopt the convention for
the Higgs doublets: $H_u  \equiv \Phi_2$,  $H_d \equiv i\tau_2 \Phi^*_1$,
where $\tau_2$  is the  usual Pauli  matrix.} and $S={1\over \sqrt{2}}
\left(v_S+\phi_S+ia_S\right)$ is the singlet Higgs field. Here
$\Delta\left[\Phi_1^0,\Phi_2^0,S\right]$ is a Coleman-Wienberg type
functional which encodes the radiative corrections. The
VEV of the effective Lagrangian $-{\mathcal L}^f_{\rm self}$ 
is equal to the fermion mass $m_f$, allowing us to substitute
for the effective Yukawa coupling $h_f$. 

We can use the self-energy effective Lagrangian 
${\mathcal L}_{\rm self}^f$ to obtain the form of the effective Lagrangian for the Higgs boson
couplings to the fermion $f$ through a Higgs boson low energy theorem
~\cite{Ellis:1975ap,BPP}.
Written in terms of the physical Higgs eigenstates $H_{1,2,3}$ and
$A_{1,2}$, the effective interaction Lagrangian is
\begin{equation}
\!
-{\mathcal L}^{\rm eff}_{\phi\bar ff} =
{g_w m_f\over 2M_W} \!\!
\left[\sum_{i=1}^3 g^S_{H_i ff} H_i\bar ff
\!+\!\!\sum_{i=1}^2 g^P_{A_i ff} A_i\!\left(\bar f i\gamma^5 f\right)
\right],
\end{equation} where the effective couplings $g^S$ and $g^P$ are
given by \cite{Hodgkinson}
\begin{equation}
\!\! g^S_{H_i ff} =
\left(1+{\sqrt{2}\over v_1}\left<\Delta_f\right>\right)^
{\!\!\!\!^{-1}}
\!\!\!\!
\left[{O^H_{1i}\over c_\beta}+\Delta_f^{\phi_2}{O^H_{2i}\over c_\beta}
+\Delta_f^{\phi_S}{O^H_{3i}\over c_\beta}\right]
\end{equation}
\begin{equation}
\!\! g^S_{A_i ff} =
\left(1+{\sqrt{2}\over v_1}\left<\Delta_f\right>\right)^
{\!\!\!\!^{-1}}
\!\!\!\!
\left[-\!\!\left(t_\beta+\Delta_f^{a_2}\right)O^A_{1i}
+\Delta_f^{a_S}{O^A_{2i}\over c_\beta}\right]
\end{equation}
Here the orthogonal matrix $O^H (O^A)$ is related to the mixing
of the CP-even (CP-odd) scalars and the loop corrections are given
by the HLET

\begin{equation}
\Delta_f^{\phi_{2,S}}=\sqrt{2}\left<
{\partial \Delta_f\over \partial \phi_{2,S}}\right>,\ \ \ 
\Delta_f^{a_{2,S}}=i\sqrt{2}\left<
{\partial \Delta_f\over \partial a_{2,S}}\right>\; .
\end{equation}
\subsection{One-loop evaluation}
\label{subsec:11}
As may be seen from the above discussion, the effective low-energy
couplings of the Higgs bosons to fermions may be calculated from the
fermion self-energies. The dominant contributions to the $b$ quark
self-energy at large $\tan\beta$ are due to squark-gluino and
squark-higgsino loops, giving

\begin{eqnarray}
\Delta_b & = &
-\; {2\alpha_s\over 3\pi}\; M_3\,
    \left(A_b\Phi^{0\ast}_1-\lambda S^\ast\Phi^{0\ast}_2\right)
I(m^2_{\tilde b_1},m^2_{\tilde b_2},M^2_3) \nonumber \\
&& +\ {{h_t^2}\over 16\pi^2}\; 
\left(A_t\Phi_2^{0\ast}-\lambda S\Phi_1^{0\ast}
\right)\nonumber\\
&& \ \times\left[\ m_{\tilde \chi_1} 
{\mathcal V}^\dag_{\{21\}}
{\mathcal U}^\ast_{\{12\}}\; 
I(m^2_{\tilde t_1}, m^2_{\tilde t_2},
 m^2_{\tilde\chi_1})\right. \nonumber \\
&&\ \ \ \ \left. +\ m_{\tilde \chi_2}
{\mathcal V}^\dag_{\{22\}}
{\mathcal U}^\ast_{\{22\}}\; 
I(m^2_{\tilde t_1}, m^2_{\tilde t_2},
 m^2_{\tilde\chi_2})\; \right]\; .
\end{eqnarray}

Here $I(a,b,c)$ is the usual $1$-loop integral function
\begin{equation}
I(a,b,c)\ =\ {{ab\ln{(a/b)}\: +\: bc\ln{(b/c)}\: +\: ac\ln{(c/a)}}\over
(a-b)(b-c)(a-c)}\; .
\end{equation}
Note that the chargino-mixing matrices
$\mathcal V,U$ are functionals of $\Phi^0_{1,2}$ and $S$, as are the
sbottom quark masses $m_{\tilde b_{1,2}}$, stop quark masses
$m_{\tilde b_{1,2}}$ and chargino masses $m_{\tilde \chi_{1,2}}$.

Similarly, the dominant $\tan\beta$ enhanced contribution to the $\tau$
lepton self-energy is due to a stau-chargino loop and is easily
derived. The effective Yukawa
couplings $\Delta_f^{\phi,a}$ are obtained as the derivatives
of these expressions. Note that the presence of the singlet in the
model does not alter the form of the $1$-loop $\tan\beta$ enhanced
couplings of the doublet Higgs fields well known from the MSSM 
\cite{CGNW}.

\vspace{-0.25in}

\section{Phenomenology}
\label{sec:3}
%\subsection{Subsection title}
%\label{sec:2} as required. Don't forget to give each section and subsection a
%unique label (see Sect.~\ref{sec:1}).
%\subsubsection{Subsubsection title} 
%\begin{table}

As the one-loop couplings of the singlet Higgs
boson to the $b$ quark and the $\tau$ lepton become significant at
large values of $\tan\beta$ and $\lambda$, we shall set $t_\beta=50$
 and $\mu={1\over\sqrt 2}\lambda v_S=110$ GeV throughout our discussion. The remaining default values of the SUSY parameters for our benchmark scenario, consistent with the constraints from LEP data, are

\begin{flushleft}
\begin{tabular}{lll}
%\hline\noalign{\smallskip}
%first & second & third   \\
$M_{\tilde Q}=300$ GeV, & $M_{\tilde L}=90$ GeV, &\\
$M_{\tilde b}=110$ GeV, & $M_{\tilde t}=600$ GeV,
& $M_{\tilde \tau}=200$ GeV,\\
$A_\tau= 1$ Tev, & $A_t= 1$ TeV, & $A_b=1$ TeV,\\
$M_1=400$ GeV, & $M_2=600$ GeV, & $M_3=400$ GeV,
\end{tabular}
\end{flushleft}
% Or use
%\vspace*{1cm}  % with the correct table height
%\end{table}

The physical Higgs boson couplings to the $b$ quark and $\tau$ lepton,
i.e. $H_{1,2,3}\bar ff$ and $A_{1,2}\bar ff$, have
contributions from both the proper vertex interaction, dominated by
the tree-level $\phi_1$ coupling, and also the mixing of the
fields $\phi_{2,S}$ with $\phi_1$. This mixing is a tree level effect and
is very significant for generic Higgs boson mass matrices. Since our interest is to assess the significance of the one loop singlet Higgs vertex effects, we focus on variants of the mnSSM and NMSSM where the mixing of $\phi_1(a_1)$ with the other scalars is suppressed.

Suppressing both the Higgs boson self-energy transitions
 $\phi_1\rightarrow\phi_{2,S}$ simultaneously is difficult,
except in the MSSM limit $\lambda\rightarrow 0$
with $\mu$ fixed, where the couplings $\Delta_f^{\phi_S(a_S)}$ also
vanish. Instead we impose a constraint on the pseudoscalar mass matrix
such that $\left(M^2_P\right)_{12}=0$. Although this condition is
arbitrarily applied here, it is robust against the dominant corrections
to the pseudoscalar mass matrix, which can absorbed into the would-be
MSSM pseudoscalar mass
$M_a$, and can be generated naturally within certain SUSY-breaking
scenarios, e.g. \cite{Dermisek:2005ar}.

\vspace{-0.15in}

\subsection{mnSSM results}
\label{sec:21}

The mnSSM is based on the renormalizable superpotential
\begin{eqnarray}
  \label{WmnSSM}
{\mathcal W}_{\rm mnSSM} &= & h_l\widehat H_d^T i\tau_2\widehat
L\widehat E\ +\ h_d\widehat H_d^T i\tau_2\widehat Q\widehat D\ \nonumber\\
&&+\; h_u\widehat Q^T i\tau_2\widehat H_u\widehat U \nonumber\\
&&+\ \lambda\widehat S\widehat H_d^T i\tau_2\widehat H_u+ t_F \widehat S
\; .
\end{eqnarray}
The term linear in $\hat S$ is induced by supergravity
quantum effects from Planck-suppressed non-renormalizable operators in
the K\"ahler  potential and superpotential \cite{comment}.

\begin{figure}
\includegraphics[width=0.45\textwidth,height=0.25\textwidth,angle=0]
{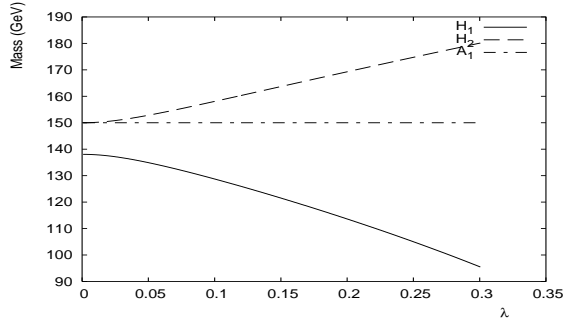}
\caption{\it Masses of the $H_1$ (solid line), $H_2$ (dashed line) and
$A_1$ (dot-dashed line) bosons in the mnSSM with $\mu=110$ GeV,
$\lambda t_S/\mu=(150$ GeV$)^2$ and $m_{12}^2=-0.5$ TeV$^2$. The
values of other soft SUSY-breaking parameters are given in 
Section~\ref{sec:3}.}
\label{fig:3}       % Give a unique label
\end{figure}
\begin{figure}
\includegraphics[width=0.45\textwidth,height=0.25\textwidth,angle=0]
{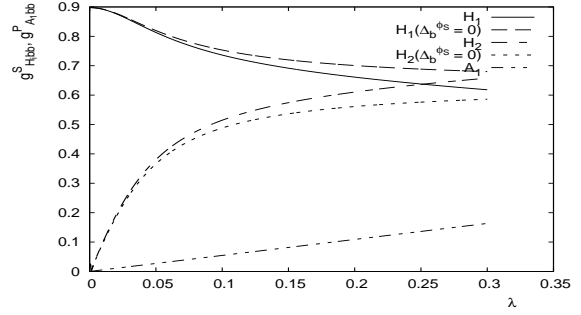}
\caption{\it  The SM-normalized  couplings $H_1b\bar  b$,
$H_2b\bar b$ and $A_1b\bar b$ in the
mnSSM, as  functions of  $\lambda$, for the  same model  parameters as
in~Fig.~\ref{fig:3}.
% Also shown are the corresponding couplings 
%$H_1b\bar b$ and $H_2b\bar b$ with the singlet threshold correction
%$\Delta^{\phi_S}_b=0$ for comparison. The pseudoscalar couplings
%are approximately zero when $\Delta^{a_S}_b=0$
.}
\label{fig:4}       % Give a unique label
\end{figure}

In Fig.~\ref{fig:3} we plot the masses of the two lightest CP-even Higgs
bosons $H_1$ and $H_2$ and the lightest CP-odd Higgs $A_1$ in the
mnSSM with $M_{H^\pm}= 5$ TeV and $\lambda t_S/\mu= (150$ GeV$)^2$. The
remaining physical Higgs states $H_3\sim\phi_1$ and $A_2\sim a$ are heavy,
of order $M_{H^\pm}$. For large values of $\lambda>0.3$ the lightest Higgs
boson mass $M_{H_1}$ is well below the LEP limit from direct Higgs
searches. Fig.~\ref{fig:4} then shows the
dependence of the $b$-quark Yukawa couplings $g^S_{H_{1,2}bb}$   and  $g^P_{A_1bb}$, for  the above  scenario.
The CP-even  Yukawa couplings $g^S_{H_{1,2}bb}$ receive
appreciable   contributions   from   the tree-level mixing  of the state
$\phi_1$ with $\phi_{2,S}$, which are competitive with the loop-induced
Yukawa couplings $\Delta^{\phi_{2,S}}_b$.   The 
coupling $g^P_{A_1 b\bar  b}\approx g^P_{a_S  b\bar b}$  is completely
dominated  by the 1-loop   contribution  $\Delta^{a_S}_b$.    For 
moderate   values  of $\lambda\sim 0.3$, we  find that 
$g^P_{A_1 b\bar b}  \sim 0.15$. Moreover,
the decay $A_1  \to \bar{b}b$ is expected to  be the dominant  decay
channel  in this  specific scenario  of the mnSSM.

\vspace{-0.15in}

\subsection{NMSSM results}
\label{sec:22}

We now  turn our  attention to the  NMSSM. The superpotential  of this
model is given by
\begin{eqnarray}
{\mathcal W}_{\rm NMSSM} &=&
h_l \widehat H_d^T i\tau_2 \widehat L\widehat E\: +\: 
h_d \widehat H_d^T i\tau_2\widehat Q\widehat D\nonumber\\
&& +\: h_u \widehat Q^T i\tau_2\widehat H_u\widehat U\nonumber\\
&&+\: \lambda \widehat S\widehat H_d^T i\tau_2\widehat H_u\: 
+\: {\kappa\over 3}\widehat S^3\; .
\end{eqnarray}

The NMSSM spectrum contains a light singlet dominated pseudoscalar
if the soft trilinear couplings are approximately $A_\lambda\sim 200$ GeV
and $A_\kappa\sim 5$ GeV. This can be naturally arranged in gauge or
gaugino mediated SUSY breaking scenarios, where these parameters are
zero at tree level. The above scales are generated by quantum corrections
if the gaugino masses are of the order $100$ GeV.

\begin{figure}
\includegraphics[width=0.45\textwidth,height=0.25\textwidth,angle=0]
{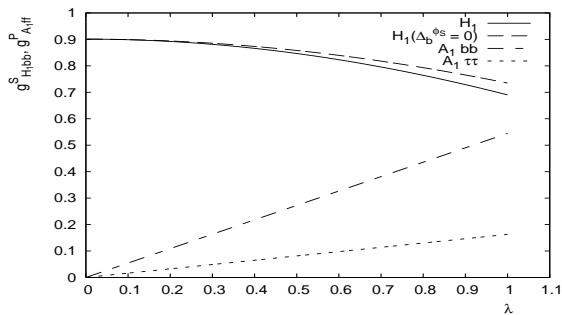}
\caption{\it The SM-normalized couplings $H_1b\bar b$~(solid
  line), $A_1b\bar b$~(dashed line)  and
  $A_1\tau^+\tau^-$~(dot-dashed line) in the NMSSM, as functions of
$\lambda$.
% We also show the coupling 
%$H_1b\bar b$ with the singlet threshold correction
%$\Delta^{\phi_s}_b=0$. Both pseudoscalar couplings $A_1ff$ are 
%approximately zero when $\Delta^{a_s}_f=0$.
}
\label{fig:5}       % Give a unique label
\end{figure}

In recent years there has been some  interest in the phenomenology
of  light Higgs pseudoscalars in the NMSSM, which may  provide an 
invisible decay
channel for a light SM-like Higgs boson.  If these CP-odd scalars have
a  large  singlet  component,  it  is  possible  for  them  to  escape
experimental  bounds  \cite{GHE}. In Fig.~\ref{fig:5} we plot the
couplings of $H_1$ to $b\bar b$ and of $A_1$ to both $b\bar b$ and
$\tau\bar\tau$ pairs for such a scenario with $M_{H^\pm}=2$ TeV.
Here $M_{A_1}$ is in the range $6\sim 9$ GeV and $M_{H_1}$ the range
$120\sim 140$ GeV.
The  threshold corrections can clearly have a significant
effect on the branching ratios of a light CP-odd singlet scalar for
moderate to large values of $\lambda$.
Previous studies have considered  detection of these particles through
decays to photon pairs  as the dominant mode \cite{Dobrescu:2000jt} in
the limit  of vanishing singlet-doublet  pseudoscalar mixing. 
Our analysis shows  that this need not be the case,  and the impact of
the hadronic decays of $A_1$ in so-called ``invisible Higgs" scenarios
should still be considered even in this limit.


%\begin{figure*}
%\includegraphics[width=1.\textwidth,height=0.34\textwidth,angle=0]{feyn2.eps}
%\hfill
%\includegraphics[width=0.45\textwidth,height=0.14\textwidth,angle=0]{susy07.eps}
%\label{fig:3}       % Give a unique label
%\end{figure*}

\section{Conclusions}

Minimal  extensions  of the  MSSM  generically  include singlet  Higgs
bosons.   Although  singlet Higgs  bosons  have  no  direct or  proper
couplings  to the SM  particles, their  interaction with  the observed
matter can still be significant as a result of two contributions.  The
first one is their mixing with  Higgs doublet states, which is
often considered in the  literature.  The second contribution is novel
and persists  even if the  Higgs doublet-singlet mixing  is completely
switched  off.  It results  from $\tan\beta$ enhanced gluino, chargino
and  squark quantum effects at the 1-loop level.

In the absence  of a Higgs doublet-singlet mixing,  the 1-loop quantum
effects we have been studying here will be the only means by which the
CP-odd singlet  may couple to  quarks and leptons. For  a sufficiently
light CP-odd singlet  scalar, with a mass below  the squark threshold,
the  loop-induced Yukawa  couplings  will provide  its dominant  decay
channel  into   $b$  quarks. This has important phenomenological
implications for studies of the NMSSM with light pseudoscalar.

%A possible  direction  for future  investigations
%will be to  calculate the off-diagonal couplings of  the singlet Higgs
%bosons  to down-type  quarks~\cite{Hiller}.  Our  effective Lagrangian
%presented here  may be  generalized to include  these flavour-changing
%neutral-current (FCNC) interactions of the Higgs bosons to quarks.  It
%would be  particularly valuable to  explore the impact of  the singlet
%Higgs-boson FCNC effects on $K$- and $B$-meson observables.

%
% BibTeX users please use
% \bibliographystyle{}
% \bibliography{}

\end{document}